\documentclass{aa}
\pdfoutput=1

\newcommand{\bol}{\boldsymbol}

\newcommand{\arcmint}{\mathrm{arcmin}}                          

\newcommand{\gt}{\ensuremath{\gamma_\mathrm{t}}}
\newcommand{\hatgt}{\ensuremath{\bar{\gamma}_\mathrm{t}}}
\newcommand{\hatgg}{\ensuremath{\bar{\gamma}_\mathrm{g}}}
\newcommand{\hatglss}{\ensuremath{\bar{\gamma}_\mathrm{LSS}}}
\newcommand{\hatgia}{\ensuremath{\bar{\gamma}_\mathrm{IA}}}
\newcommand{\tildegt}{\ensuremath{\hat{\gamma}_\mathrm{t}}}

\newcommand{\chid}{\ensuremath{\chi_\mathrm{d}}}
\newcommand{\Ds}{\ensuremath{D_\mathrm{s}}}
\newcommand{\Dds}{\ensuremath{D_\mathrm{ds}}}
\newcommand{\Dd}{\ensuremath{D_\mathrm{d}}}
\newcommand{\zs}{\ensuremath{z_\mathrm{s}}}
\newcommand{\zd}{\ensuremath{z_\mathrm{d}}}

\newcommand{\wdI}{\ensuremath{w_\mathrm{dI}}}
\newcommand{\wdd}{\ensuremath{w_\mathrm{dd}}}

\usepackage{amsmath, amssymb}
\usepackage[english=british]{csquotes}
\usepackage[T1]{fontenc}
\usepackage[utf8]{inputenc}
\usepackage{lmodern}
\usepackage{graphicx}
\usepackage{txfonts}
\usepackage{natbib}
 \bibpunct{(}{)}{;}{a}{}{,}
\usepackage{subfigure}
\usepackage{hyperref}

\makeatletter
\renewcommand*\aa@pageof{, page \thepage{} of \pageref*{LastPage}}
\makeatother
\begin{document} 
\title{Magnification bias in the shear-ratio test:\\a viable mitigation strategy}
\author{Sandra Unruh\inst{1}, Peter Schneider\inst{1} \and Stefan Hilbert\inst{2}}
\institute{
    Argelander-Institut f\"ur Astronomie, Universit\"at	Bonn, Auf dem H\"ugel 71, D-53121 Bonn, Germany\\	sandra, peter@astro.uni-bonn.de
    \and
    Exzellenzcluster Universe, Boltzmannstr. 2, D-85748 Garching, Germany\\
    Ludwig-Maximilians-Universit{\"a}t, Universit{\"a}ts-Sternwarte, Scheinerstr. 1, D-81679 M{\"u}nchen, Germany
}
\date{Version \today; received August 28, 2018, accepted January 17, 2019} % day month year 1 January 2005
\abstract{
Using the same lens galaxies, the ratios of tangential shears for different source galaxy redshifts is equal to the ratios of their corresponding angular-diameter distances. This is the so-called shear-ratio test (SRT) and it is valid when effects induced by the intervening large-scale structure (LSS) can be neglected. The dominant LSS effect is magnification bias which, on the one hand, induces an additional shear, and on the other hand, causes a magnification of the lens population. Our objective is to quantify the magnification bias for the SRT and show an easy-to-apply mitigation strategy that does not rely on additional observations. We use ray-tracing data through the Millennium simulation to measure the influence of magnification on the SRT and test our mitigation strategy. Using the SRT as a null-test we find deviations from zero up to \(10 \%\) for a flux-limited sample of lens galaxies, which is a strong function of lens redshift and the lens-source line-of-sight separation. Using our mitigation strategy we can improve the null-test by a factor of \(\sim \!100\).
}
\keywords{cosmological parameters -- gravitational lensing: weak, magnification -- large-scale structure}
\titlerunning{Magnification bias in the shear-ratio test} 
\maketitle
%
%
%________________________________________________________________
%
%
%
\section{\label{Sc1}Introduction}
Light bundles from background galaxy images get coherently distorted as they travel through the inhomogeneous Universe. We refer to that effect as gravitational lensing. The distortion includes a change of the intrinsic shape of the galaxies (lensing shear) as well as a magnification effect which affects the observed number density of galaxies. In the regime of weak gravitational lensing the change of galaxy shapes cannot be seen for single galaxies. A statistical approach is needed where information is extracted from thousands to millions of lens galaxies. Amplitude and direction of the distortion depend on the integrated tidal gravitational field along the line-of-sight as well as the curvature of the Universe, which makes weak lensing measurements a powerful cosmological probe \citep[see, e.g.,][for a recent review]{kilbinger2015}. Arguably, it is the most powerful method to constrain the equation of state of Dark Energy \citep{albrecht2006}. Ongoing surveys like the Hyper Suprime-Cam SSP Survey \citep{aihara2017}, KiDS \citep[Kilo Degree Survey,][]{dejong2017} and DES \citep{abbott2016} already put tight constraints on cosmological parameters \citep{troxel2017, hildebrandt2017} especially in combination with other probes \citep{vanuitert2018, joudaki2017, harnoisderaps2017}. In the future even bigger projects are planned with the Euclid mission \citep{laureijs2011}, the Square Kilometer Array \citep[SKA,][]{blake2004}, and the Large Synoptic Survey Telescope \citep[LSST,][]{ivezic2008}.

The correlation between the positions of a foreground lens population and the shear of more distant background source galaxies has been named galaxy-galaxy lensing \citep[GGL; see, e.g.,][]{hoekstra2013}. The excess projected mass around lens galaxies within an aperture \citep{squires1996, schneider1996} is reflected by the so-called tangential shear \gt. %Its most common application is mapping the matter distribution around lens galaxies with no distinction of the type of matter.
In 2003, Jain \& Taylor proposed the shear-ratio test (SRT) as a purely geometrical probe for cosmology. If the maximum separation over which galaxies are correlated with the large-scale structure (LSS) is small compared to the angular-diameter distance between lens and source, the ratio of tangential shear values from two different source populations with the same lens population does only depend on a ratio of angular-diameter distances. Hence, there is no contribution from the lens properties anymore, while the distance ratios depend on cosmology through the distance-redshift relation. Therefore, shear-ratios were originally constructed for probing cosmological parameters \citep{jain2003, bernstein2004}. However, as it turns out, the dependence on cosmology is rather weak \citep[][Zhao \& Schneider in prep.]{taylor2007} and correspondingly, first applications of this shear-ratio test returned only weak constraints \citep{kitching2007, taylor2012}. Alternative probes, e.g., CMB \citep{ade2016}, Supernovae Type 1a \citep{scolnic2018}, and cosmic shear \citep{hildebrandt2017}, provide far more accurate constraints on cosmological parameters, which essentially renders the SRT non-competitive for its original purpose. Yet, we can turn the argument around and use the SRT as a null test to detect remaining systematics \citep[see, e.g.,][]{schneider2016, prat2017}.

The main challenge in ongoing and future weak lensing surveys is to obtain an unbiased estimate of shear from faint background galaxies and their corresponding redshift distributions. Great efforts to understand what influences/biases the data beside shape distortion by the weak lensing effect \citep[see, e.g.,][]{mandelbaum2017, harnoisderaps2017, amon2017, zuntz2017, abbott2017} uncovered, e.g., noise bias \citep{bartelmann2012, melchior2012}, underfitting bias \citep{voigt2010}, and intrinsic alignment effects \citep{troxel2015}.  Moreover, photometric redshift estimates suffer from dust obscuration in galaxies, the lack of a sufficient number of spectroscopic galaxy redshifts for the calibration, and a limited number of spectral bands and galaxy template spectra. Since biases in the data, if uncorrected, can be a magnitude larger than the effects from weak lensing, different strategies including null tests have been proposed to check for remaining systematics. Commonly used null tests are correlations between corrected galaxy shapes and uncorrected stellar ellipticity \citep[e.g.,][]{heymans2012} or other quantities that should be independent of shear, as well as checking for B-mode statistics  with the cross shear or using COSEBIs \citep[Complete Orthogonal Sets of E-/B-mode Integrals,][]{schneider2010, asgari2017}. The SRT emerged among other more recently introduced probes for systematics \citep[see, e.g.,][for alternative probes]{cai2016, li2017}.

Magnification bias in GGL and the SRT has been considered by \citet[][]{ziour2008}. \citet{hilbert2009} showed that magnification bias suppresses the GGL signal expected from shear that is induced by matter correlated with the lens galaxies, by up to $20\%$. For galaxies in the Canada-France-Hawaii Telescope Lensing Survey (CFHTLenS) the magnification bias has a confirmed impact of $\sim 5\%$ \citep{simon2018}. The LSS between us and the lens will shear source galaxies. While this shear is not correlated with the true positions of the lens galaxies, magnification by the LSS also affects the observed number density of lenses and thus leads to a correlation between observed lens positions and source shear \citep{bartelmann2001}. The induced correlation between foreground and background galaxy populations has already been measured in different surveys \citep[see, e.g.,][]{scranton2005, garcia2018}. Despite the fairly large effect of this magnification bias on the GGL signal shown by \citet{hilbert2009}, it appears to have been neglected in (almost) all observational studies of GGL and their quantitative interpretation.

In this paper we will investigate the effect of magnification bias on the SRT. Since magnification bias is a function of source and lens redshift, the SRT can fail even if shear and redshift data is sound \citep{ziour2008}. We use simulated data to quantify the magnification bias in the SRT, describe its properties and, most importantly, we will present a simple mitigation strategy. The main advantage of this mitigation method is that no additional measurements are needed. We also consider a second mitigation strategy that employs stellar velocity dispersion measurements to estimate the magnification.

This paper is organized as follows: in Sect.~\ref{Sc2} we will revisit the basics of the SRT in more detail. In Sect.~\ref{Sc3} we will briefly describe our synthetic lensing data taken from ray-tracing through the Millennium simulation as well as the mock lens catalogue. The effects of the magnification bias and its mitigation strategies will be described in Sect.~\ref{Sc4} and Sect.~\ref{Sc5}. We will conclude in Sect.~\ref{Sc6}.
\section{\label{Sc2}Weak lensing and the shear-ratio test}
\subsection{Cosmological distances}
The comoving distance \(\chi(z_1, z_2)\) of a source at redshift \(z_2\) from a lens at redshift \(z_1\) is given by an integration over the Hubble parameter \(H(z)\) as a function of redshift \(z\)
\begin{align}
    \label{eq:distredshift}
    \chi(z_1, z_2) &= \int_{z_1}^{z_2} \frac{c\,\mathrm{d} z'}{H(z')} \;,
\end{align}
where
\begin{align}
    \left(\frac{H(z)}{H_0} \right)^2 &= \Omega_\mathrm{m}(1+z)^3 + (1-\Omega_\mathrm{m}) \;,
\end{align}
for a flat universe with matter density \(\Omega_\mathrm{m}\) in units of the present-day critical density \(\rho_{\mathrm{crit},0} = 3 H_0^2 / (8\pi G)\). The comoving distance \(\chi\) is related to the angular-diameter distance \(D\) via:
\begin{equation}
    D(z_1, z_2) = \frac{\chi(z_1, z_2)}{1+z_2} \;.
\end{equation}
Note that \(\chi(z_1, z_2) = \chi(0,z_2) - \chi(0,z_1)\) but \(D(z_1, z_2) \neq D(0,z_2) - D(0,z_1)\) except for \(z_2 - z_1 \ll 1\). In the following we will omit the argument zero in the distances, i.e.~\(D(0,z) := D(z)\).
\subsection{Galaxy-galaxy lensing}
Foreground matter at a redshift \zd\ will deflect light rays from background galaxies and induce a shear pattern. In complex notation the shear reads
\begin{equation}
    \gamma (\bol{\theta}) = \gamma_1 (\bol{\theta}) + \mathrm{i} \gamma_2 (\bol{\theta}) \;.
\end{equation}
Here, \(\bol{\theta}\) is the position on the sky and \(\gamma_{1,2}\) are the Cartesian shear components at angular position \(\bol{\theta}\). In GGL, the shear is measured with respect to the connecting line between a lens at position \(\bol{\theta}_\mathrm{d}\) and a source galaxy -- orthogonal to that line is the tangential shear \gt, and the cross shear \(\gamma_\times\) is measured with a \(45^\circ\)-rotation. For a fixed lens position \(\bol{\theta}_\mathrm{d}\) this corresponds to a rotation of the shear components
\begin{equation}
    \label{eq:tangshear}
    \gt (\bol{\theta}) + \mathrm{i} \gamma_\times(\bol{\theta}) = -\gamma (\bol{\theta}) \, \frac{(\bol{\theta} - \bol{\theta}_\mathrm{d})^*}{\bol{\theta} - \bol{\theta}_\mathrm{d}} \;,
\end{equation}
where we also conveniently write the position on the sky in complex notation, i.e.~\(\bol{\theta} = \theta_1 + \mathrm{i} \theta_2\), and an asterix denotes complex conjugation.

Shear is caused by a foreground line-of-sight over-density. The lensing strength factorizes into a part containing all the lens properties, and one containing the angular-diameter distances between us and source, \(D(\zs)=\Ds\), as well as between lens and source, \(D(\zd, \zs) = \Dds\) \citep[see, e.g.,][]{schneider1992}. The lens properties are characterized by its matter distribution and its angular-diameter distance, \Dd . Shear that is caused by the matter associated with the lens galaxies will be denoted with \(\gamma_\mathrm{g}\), where the `g' refers to galaxy. The expectation value of this tangential shear measurement can be written as
\begin{align}
    \label{eq:shear_exp_value}
    \gamma_\mathrm{g} (\bol{\theta}; \zd,\zs) = \gamma_{\mathrm{g},\infty} (\bol{\theta}; \zd) \, \frac{\Dds}{\Ds} := \gamma_{\mathrm{g},\infty}  (\bol{\theta}; \zd) \, \beta(\zd,\zs) \;,
\end{align}
where \zd\ is the redshift of the lens for background sources at redshift \zs . The lensing efficiency \(\beta\) is a ratio of angular-diameter distances which is scaled by \(\gamma_{\mathrm{g},\infty}\). If not noted otherwise further expressions of shear in this paper are tangential shear estimates.
\subsection{The classical shear-ratio test}
We can calculate a weighted integral to obtain a mean shear estimate \hatgg
\begin{align}
    \label{eq:galshear}
    \hatgg (\bol{\theta}) &= \int \mathrm{d}^2 \theta' \, \gamma_\mathrm{g}(\bol{\theta}+\bol{\theta}') w(|\bol{\theta}'|) \;,
\end{align}
where
\begin{align}
    w(\theta) = \frac{1}{2\pi \theta^2} \, \mathcal{H}(\theta - \theta_\mathrm{in}) \, \mathcal{H}(\theta - \theta_\mathrm{out}) \;,
\end{align}
is a weight function different from zero only in the annulus \(\theta_\mathrm{in} \leq \theta \leq \theta_\mathrm{out}\) and where \(\mathcal{H}\) is the Heaviside step function. This form of the weight function optimizes the signal-to-noise ratio for a shear profile behaving like \(1/\theta\), as is the case for an isothermal profile \citep{bartelmann2001}.

If the tangential shear can be factorized as in Eq.~\eqref{eq:shear_exp_value}, i.e.~into a factor that depends only on lens properties and the lensing efficiency, we can consider shear measurements from two different source populations at \(z_i\) and \(z_j\) behind the same lens galaxy or a population of lens galaxies at fixed redshift \zd. Then the ratio, \(R\), of those shear measurements is independent of the lens properties and is solely determined by the geometry of the observer-lens-source system as a ratio of lensing efficiencies,

\begin{align}
    \label{eq:SRT}
     \frac{\hatgg (\zd, z_j)}{\hatgg ( \zd, z_i)} = \frac{\beta(\zd,z_j)}{\beta(\zd,z_i)} =:  R(\zd; z_i, z_j) \;.
\end{align}
Eq.~\eqref{eq:SRT} can be written as a null test which can be applied in cosmic shear measurements as a consistency check that does not require any additional data. This method is purely based on geometrical considerations and as such it is independent of structure growth in the universe. This makes the SRT easy-to-apply since it does not require the use of simulations.
\section{\label{Sc3}Mock data}
\subsection{Millennium simulation data}
In this work we make use of the Millennium simulation \citep[MS,][]{springel2005}. The MS is an N-body simulation tracing the evolution of \(2160^3\) dark matter particles of mass \(8.6 \times 10^8 \, h^{-1} \mathrm{M}_\odot\) enclosed in a \(\left(500 \, h^{-1} \mathrm{Mpc}\right)^3 \)-cube, where \(h\) is the dimensionless Hubble parameter defined as \(H_0 = 100 h \, \mathrm{km \, s^{-1} \, Mpc^{-1}}\). In the MS, 64 snapshots are available in the redshift range from \(z=127\) to today. The underlying cosmology is a flat \(\Lambda\)CDM cosmology with matter density parameter \(\Omega_\mathrm{m} = 0.25\), baryon density parameter \(\Omega_\mathrm{b} = 0.045\), dark energy density parameter \(\Omega_\Lambda = 1 - \Omega_\mathrm{m} = 0.75\), a dimensionless Hubble parameter \(h = 0.73\), a scalar spectral index \(n_\mathrm{s} = 1\) and a power spectrum normalization of \(\sigma_8 = 0.9\). These values agree with a combined analysis of 2dFGRS \citep{colless2001} and first-year WMAP data \citep{spergel2003}.

Various catalogues of galaxies have been added to the simulation using semi-analytic galaxy-formation models. \citet{saghiha2017} showed that the galaxy catalogue from the model by \citet{henriques2015} matches best with the observed GGL and galaxy-galaxy-galaxy lensing signal from the CFHTLenS. We use lens galaxies from the redshift slices 59 to 43 corresponding to \(z_{59} = 0.0893\) and \(z_{43} = 0.8277\).

Furthermore, we use simulated lensing data obtained by a multiple-lens-plane ray-tracing algorithm in 64 realizations with a \(4 \times 4 \, \mathrm{deg}^2\)-field-of-view \citep{hilbert2009}, where we consider the source redshift planes 58 to 34 that correspond to redshifts \(z_{58} = 0.1159\) and \(z_{34} = 1.9126\). To perform the ray-tracing, the matter distribution in each redshift slice is carefully mapped to the mid plane. Then, a multiple lens plane algorithm is used to calculate shear, magnification and convergence information on a grid in each mid plane. With a box size of \(500 \, h^{-1} \mathrm{Mpc}\), the MS is not large enough to contain a full light cone out to high redshifts. However, a simple stacking of simulation cubes would result in a light ray that meets the same matter structures several times on its way to \(z=0\) due to periodicity. Therefore, \citet{hilbert2009} decided to perform the ray-tracing using a skewed angle through the box, yet making use of its periodic boundary conditions. Hence, no random rotation or translation of the matter in the box has been done, which preserves the galaxy-matter correlation. Effectively, a light ray can travel a comoving distance of \(5 \, h^{-1} \mathrm{Gpc}\) before encountering the same matter structures.

To avoid double counting, the indices in Eq.~(\ref{eq:SRT}) will be restricted to \(i<j\) since \(R(\zd; z_i, z_j) = 1/R(\zd; z_j, z_i)\). We will also only use consecutive redshift bins for taking ratios due to the relation \(R(\zd;z_i,z_k) = R(\zd;z_i,z_j)\,R(\zd;z_j,z_k)\).
\subsection{Obtaining a tangential shear estimate}
We can calculate the weighted mean tangential shear
\begin{align}
    \label{eq:shearestimator}
    \hatgt (\bol{\theta}) &= - \Re \left[ \int \mathrm{d}^2 \theta' \, \gamma (\bol{\theta}+\bol{\theta}') \, \frac{\bol{\theta}'^*}{\bol{\theta}'} w(|\bol{\theta}'|) \right] \;,
\end{align}
for every grid point of the \(4 \times 4 \, \mathrm{deg}^2\)-field. It defines a convolution which reduces to a simple multiplication in Fourier space. We thus use Fast Fourier transforms (FFT) from the library of \citet{FFTW} to compute the shear estimator~\eqref{eq:shearestimator}. A fast Fourier Transform (FFT) implicitly assumes periodic boundary conditions. Since we convolve the shear field with a function of finite support, results from the FFT will be wrong in a stripe of the thickness \(\theta_\mathrm{out}\) around the field edge. Thus, we do not consider shear data in this stripe and focus on the data in the inner \((4^\circ -2\theta_\mathrm{out}) \times (4^\circ -2\theta_\mathrm{out})\) area of the field. For the inner boundary of the annulus we will set \(\theta_\mathrm{min} = 0.\!'5\) in the paper if not explicitly noted otherwise. The exact value of \(\theta_\mathrm{min}\) is not crucial, since the weighting function will give the same shear signal per logarithmic bin if the mass profile is isothermal. However, we cannot go to arbitrarily small \(\theta_\mathrm{in}\) due to the finite resolution of the simulation.

We extract the positions of galaxies from the \citet{henriques2015} model catalogue that fulfill our selection criteria, e.g., a flux limit or a cut in halo mass. The positions of the \(N_\mathrm{L}\) galaxies are then assigned to their nearest grid point. Each pixel in the grid has a size of \((3.5\,\mathrm{arcsec})^2\). Thus, as long as we choose the size of our integration area in (\ref{eq:shearestimator}) in the square arcminute regime, we will not suffer from discretization effects. We then simply average the shear signal (\ref{eq:shearestimator}) for all lens galaxies
\begin{align}
    \label{eq:shearest}
    \langle \hatgt \rangle &= \frac{1}{N_\mathrm{L}} \sum_{i=1}^{N_\mathrm{L}} \hatgt (\bol{\theta}_i) \;.
\end{align}

When we apply a cut at \(24\,\mathrm{mag}\) in the \(r\)-band, we typically find 1500 lens galaxies per field in the lowest redshift slice and around 30\,000 lens galaxies in the highest redshift slice. To reduce the statistical error we use the shear information on all \(4096^2\) pixel (in contrast to use shear information only on those positions where a source galaxy is located according to the Henriques galaxy catalogue).

We repeat the whole process but this time we randomize the lens galaxy positions. \citet{singh2017} showed that subtracting the signal around random points from the actual shear signal usually leads to a more optimal estimator with a decreased error budget. Although this poses only a minor contribution in our case, we change our shear estimator to \(\hatgt \to \hatgt - \bar{\gamma}_\mathrm{t, rand}\).
\section{\label{Sc4}The magnification bias}
\subsection{The conventional shear-ratio test}
\begin{figure*}[htbp]
    \centering
    \includegraphics[width=\textwidth]{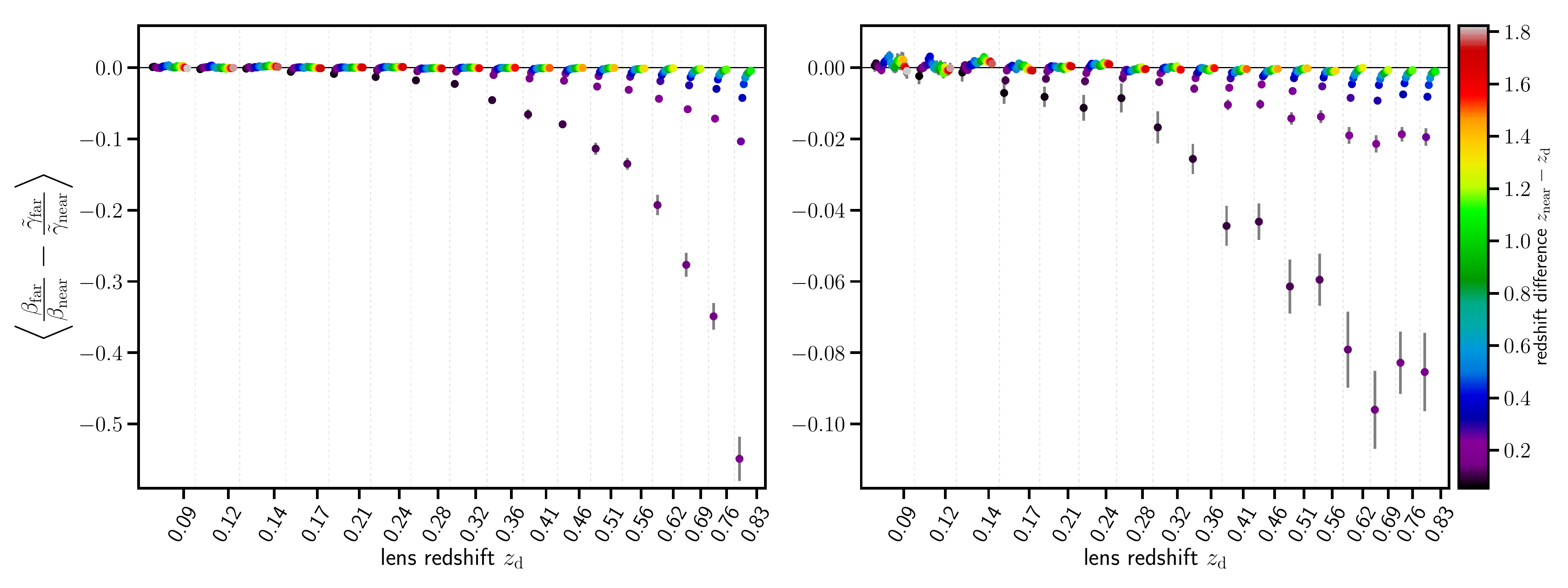}
    \caption{Shear-ratio test of two populations of galaxies that are sheared by the same lens, i.e. \(z_\mathrm{far} > z_\mathrm{near} > \zd\) with source redshifts in consecutive bins. The dotted lines separate shear-ratios from different lens galaxies, while source redshifts increase from left to right which is highlighted by color code. Each color represents a combination of two sources at \(z_\mathrm{near}\) and \(z_\mathrm{far}\). Ideally the outcome of the SRT (\ref{eq:nullhyp}) is zero. Data is taken from ray-tracing through the Millennium simulation. The shear estimator \(\langle\hatgt\rangle\) is defined in Eq.~(\ref{eq:shearest}) with \(\theta_\mathrm{in} = 0.\!'5\) and \(\theta_\mathrm{out} = 5'\). Errors are obtained by a Jackknife method from the 64 different realizations per redshift. \textbf{Left}: all simulated galaxies have been used. \textbf{Right}: a magnitude cut of \(24\,\mathrm{mag}\) in the \(r\)-band is applied for the lens galaxies.}
    \label{pic:SRT}
\end{figure*}

We calculate ratios of observed shear estimates and consider the null-test
\begin{align}
    \label{eq:nullhyp}
    \left\langle\frac{ \hatgg (\zd, z_j)}{\hatgg ( \zd, z_i) }\right\rangle - \frac{\beta(\zd,z_j)}{\beta(\zd,z_i)} = 0\;.
\end{align}
We use the shear-ratios for all statistically independent redshift bins and show the result of the SRT in Fig.~\ref{pic:SRT}. Since our method is unaffected by shape noise and Poisson noise of source galaxies, we expect
only minor deviations from zero in the SRT. However, this is not the case. The deviation from zero gets worse for higher lens redshift and smaller separations between lens and sources. In the right figure we applied a cut in magnitude which includes the magnification (and is thus the observed magnitude), in contrast to the right figure where we used all available galaxies in the catalogue which is equivalent to a stellar mass-limited sample. In the lowest-redshift bins (i.e., \(\zd<0.15\)) the results are almost identical since \(>95\%\) of the lenses are brighter than \(24\,\mathrm{mag}\) in the \(r\)-band. For these redshift bins, the classical SRT (\ref{eq:nullhyp}) performs as expected. In the medium-redshift range \(0.15 \leq \zd \leq 0.4\) a sensible choice of lens and source redshifts will keep the deviations small. The effect for the magnitude-limited sample is quantitatively smaller but qualitatively similar. The bias is strongest for high-redshift lenses with \(\zd > 0.4\) where even widely separated lenses and sources show deviations from zero at the percent level.
\subsection{Magnification effects}
Until now, we have only considered the shear caused by the matter that is associated with the lens galaxies at fixed redshift \zd. However, there exists intervening LSS between us and the source that can induce an additional shear signal. This shear is not correlated with the true positions of lens galaxies. However, the LSS also alters the distribution of lens galaxies on the sky by magnification effects. This leads to a correlation between shear caused by the LSS and the \textit{observed} distribution of galaxies. For most practical purposes magnification bias is the dominant second-order effect \citep{hui2007}.

We can approximate the influence of magnification by the LSS with
\begin{align}
    \hatgt (\zd, \zs) &= \hatgg (\zd, \zs) + \hatglss (\zs) \;.
\end{align}
The shear at the source redshift is a superposition of the shear induced by matter associated with the lens galaxies at redshift \zd\ and the LSS between us and the source galaxies. An average of the shear contribution from the LSS over a sufficient number of source galaxy must be zero. However, we use lens galaxies to obtain an averaged tangential shear estimate and the observed position of these galaxies is altered by the intervening LSS. This leads to a correlation between LSS-induced shear and lens galaxy positions:
\begin{align}
    \label{eq:shearobs}
    \langle \hatgt \rangle (\zd, \zs) &= \langle \hatgg \rangle (\zd, \zs) + \langle \hatglss \rangle (\zd, \zs) \;.
\end{align}

\begin{figure}[htbp]
    \centering
    \includegraphics[width=.49\textwidth]{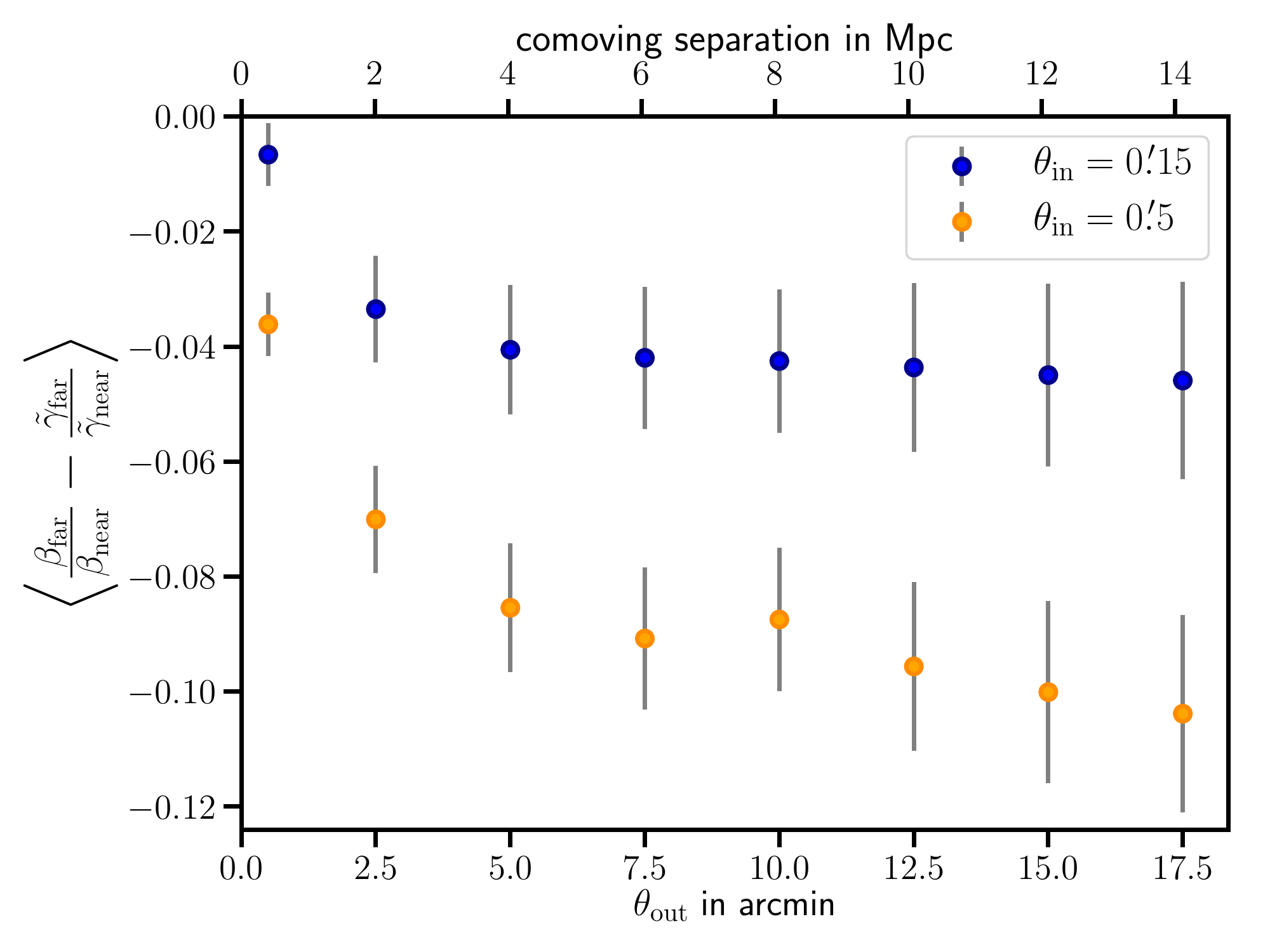}
    \caption{For an identical setup as the right figure~\ref{pic:SRT} we perform the SRT; we vary the integration range by choosing two different \(\theta_\mathrm{in}\) and by altering \(\theta_\mathrm{out}\) in the shear estimator \eqref{eq:shearestimator} for each \(\theta_\mathrm{in}\). The leftmost points are shear estimates around a thin annulus at \(\theta=0.\!'15\) (blue) and \(\theta=0.\!'5\) (orange). For clarity only the outcome of the SRT for the combination \(\zd=0.83\), \(z_\mathrm{near} =0.91\) and \(z_\mathrm{far} =0.99\) is shown. The SRT differs strongly from zero regardless of \(\theta_\mathrm{out}\), only at a thin annulus very close to the lens is the signal almost compatible with zero.}
    \label{pic:devSRT}
\end{figure}

The relative contribution of the foreground LSS to the observed lensing signal depends on the size of the annulus. We expect that for a small annulus close to the lens position, the mean lensing signal is dominated by shear associated with matter at the lens redshift \zd. Hence, we investigate the impact of \(\theta_\mathrm{out}\) for two different \(\theta_\mathrm{in}\) for the shear estimator \hatgt. We concentrate on the realistic case of a flux-limited sample and for each \(\theta_\mathrm{in}\) we vary \(\theta_\mathrm{out}\) for an SRT as done in Fig.~\ref{pic:SRT}. For clarity we only plot the SRT for the lens-source-source combination \(\zd = 0.83\), \(z_\mathrm{near} = 0.91\) and \(z_\mathrm{far} = 0.99\) which corresponds to one the largest deviations from zero in our SRT. The result shown in Fig.~\ref{pic:devSRT} follows our expectations. The deviation 
from zero is less pronounced for small annuli but it is still present. For larger integration ranges, the deviation is larger, but stays approximately constant for \(\theta_\mathrm{out} \geq 5'\).

We verify that it is indeed magnification that affects the shear estimate around galaxies. Since the influence grows with redshift, with smaller redshift differences of lens and source galaxies (Fig.~\ref{pic:SRT}), and with the size of the annulus \citep[][see Fig.~\ref{pic:devSRT}]{ziour2008}, we choose (again) our highest redshift bins with \(\zd = 0.83\) and the two consecutive redshift bins as source galaxies with \(z_\mathrm{near} = 0.91\) and \(z_\mathrm{far} = 0.99\). We set the integration range to \(\theta_\mathrm{out} = 17.\!'5\). We bin the lens galaxies in magnification such that each bin contains a roughly equal number of lenses. Then, we measure the shear in the two source planes, \(\langle \hatgt \rangle (\zd, z_\mathrm{near})\) and \(\langle \hatgt \rangle (\zd, z_\mathrm{far})\), for each bin and plot it against the average magnification per bin. Furthermore, we want to visualize the influence the LSS has on the measured shear result. Since we chose lens and source plane close to each other, almost all the relevant LSS is also in front of the lenses. Then, we can just measure the shear signal \(\langle \hatgt \rangle (\zd, \zd)\) around the lens galaxies at the lens redshift to get an approximate measurement of \(\langle\hatglss\rangle(\zd,\zs)\).

\begin{figure}[htbp]
    \centering
	\includegraphics[width=.49\textwidth]{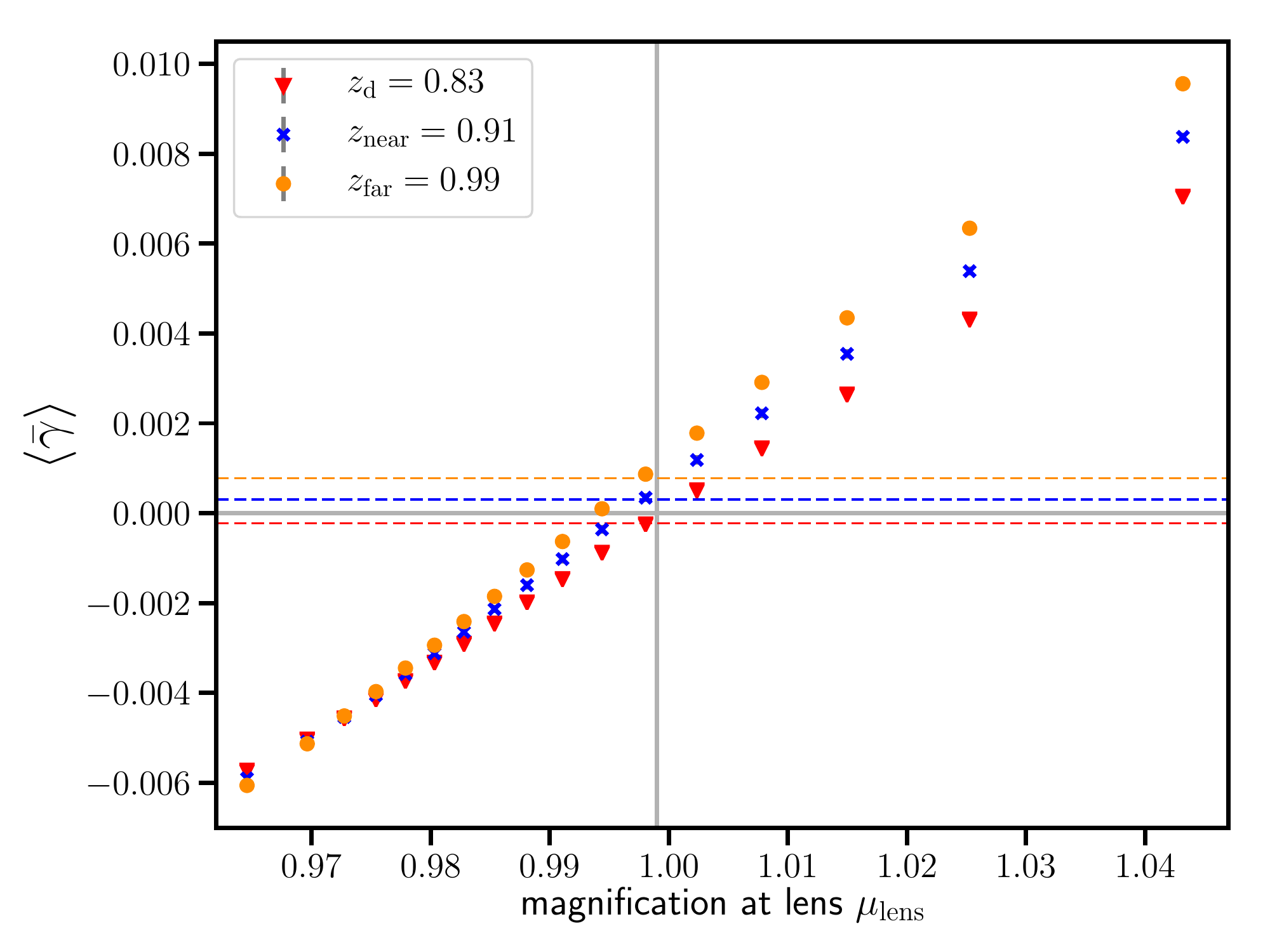}
	\caption{The shear estimate \(\langle \hatgt \rangle (\zd=0.83, z_i)\) with \(\theta_\mathrm{in} =  0.\!'5\) and \(\theta_\mathrm{out} =  17.\!'5\) is shown. A grey vertical line indicates the mean magnification of all lens galaxies. The lens galaxies have been binned in magnification with a roughly equal number of lenses in each bin. We omitted the two highest bins in magnification at mean magnification \(\mu_\mathrm{lens} = 1.11\) and \(\mu_\mathrm{lens} = 1.6\) for clarity. The dashed lines represent the mean shear estimates for all lenses. Due to the magnification effects of the intervening LSS the shear around lens galaxies at lens redshift (red triangles) differs from zero as does the red dashed line which is an approximate measure for \hatglss \ in Eq.~(\ref{eq:shearobs}). A clear correlation between shear and magnification can be seen.}
	\label{pic:shearmagn}
\end{figure}

The result is displayed in Fig.~\ref{pic:shearmagn}, 
where we also plot the mean shear for all lens galaxies irrespective of magnification as dashed lines. As expected from the form of the lensing kernel, the shear from sources with higher redshift \(\langle \hatgt \rangle (\zd, z_\mathrm{far})\) is larger than \(\langle \hatgt \rangle (\zd, z_\mathrm{near})\). Shear and magnification show a clear correlation. The red dashed line, however, is naively expected to be consistent with zero, as it would be the case if all the contributions to the lensing signal came from matter associated with the lens galaxies. However, there exists an additional contribution from shear caused by the LSS in front of the lens galaxies. Due to the selection of lens galaxies in the foreground (in contrast to random positions) the LSS-induced shear does not vanish. It can be seen that the red dashed line, a measure for \hatglss, is of the same order of magnitude as the blue dashed line, the shear signal for sources \(\zs>\zd\).
\section{\label{Sc5}Mitigation strategies}
As we have seen in the previous section, the influence of the magnification bias is negligible for low lens redshifts as well as for lenses and sources separated widely in redshift. This was already pointed out in \citet{moessner1998}. However, as we approach Stage IV experiments to infer the equation of state parameter for Dark Energy, we go to higher and higher lens and source redshifts. A mitigation strategy is thus crucial. Using theoretical considerations for lensing power spectra, \citet{ziour2008} derived a mitigation in their Eq.~(38) for the SRT. It involves the knowledge of the easily obtainable number count slope as well as the linear galaxy bias factor at of lenses as a function of mass, which is notoriously difficult to obtain. The assumption of a linear bias factor will eventually break down for small angular scales, and even on large scales, a linear bias is not necessarily a sufficiently accurate approximation \citep{hui2007}. 

In the following, we will introduce a new mitigation strategy. Its main advantage is that it does not require additional observations or simulations. As can be seen in Fig.~\ref{pic:shearmagn}, the shear signal \(\langle \hatgt \rangle(\zd, \zd)\) is not zero if measured at the lens redshift. However, the shear induced by matter associated with the lens galaxy \(\langle \hatgg \rangle(\zd, \zd)\) is certainly zero. Thus, what we measure is due to the intervening LSS
\begin{equation}
    \label{eq:estimator_for_hatglss_of_z_d_z_d}
    \langle \hatgt \rangle(\zd, \zd)    = \langle \hatglss \rangle(\zd,\zd). 
\end{equation}

In general, the influence of the LSS grows with redshift, and if the separation of lenses and sources is moderate, we can introduce a scaling factor \(\lambda \gtrsim 1\) that parametrizes this similarity as \(\langle \hatglss \rangle(\zd,\zs) \approx \lambda \langle \hatglss \rangle(\zd,\zd)\). From the form of the lensing kernel, we can deduce that the main contribution of the LSS to the shear signal is located at about half the distance between us and the source. Therefore, if we increase the source redshift only slightly, we will also increase \(\lambda\) slightly. Using these assumptions it follows naturally that the SRT performs better for high source redshifts at fixed lens redshift. Relative to the shear contribution from the LSS, the shear from matter associated with the lens galaxies shows a strong dependence on source distance. Thus, the LSS-induced shear has less impact for larger line-of-sight separations of lens and sources.

\subsection{Improved SRT -- quantifying the foreground contribution}

The scaling factor can be obtained from the data by correlating the observed lensing signal $\langle \hatgt \rangle (\zd, \zs) $ at the source redshift with the foreground part \(\langle \hatglss \rangle (\zd, \zd)\). Using Eq.~\eqref{eq:shearobs}, we find
\begin{align}
    \langle \hatgt (\zd,\zs) \hatglss(\zd,\zd) \rangle &= \langle \hatgg (\zd,\zs) \hatglss (\zd,\zd) \rangle \nonumber\\
    &+ \lambda (\zd,\zs) \, \langle \hatglss(\zd,\zd) \hatglss(\zd,\zd) \rangle \;,
\end{align}
where averages are taken as in Eq.~\eqref{eq:shearest} for each lens. The first term on the right-hand side, \(\langle \hatgg \hatglss \rangle\), vanishes since the shear from matter associated with the lens galaxies is uncorrelated with the shear caused by the foreground matter. Exploiting Eq.~\eqref{eq:estimator_for_hatglss_of_z_d_z_d}, we can calculate the scaling factor by
\begin{align}
\label{eq:fudgefac}
    \lambda (\zd,\zs) = \frac{\langle \hatgt (\zd,\zs) \hatglss (\zd,\zd) \rangle}{\langle \hatglss (\zd,\zd) \hatglss (\zd,\zd) \rangle} 
        =
        \frac{\langle \hatgt (\zd,\zs) \hatgt (\zd,\zd) \rangle}{\langle \hatgt (\zd,\zd) \hatgt (\zd,\zd) \rangle} \;.
\end{align}

\begin{figure*}[htbp]
    \sidecaption
    \centering
    \includegraphics[width=12cm]{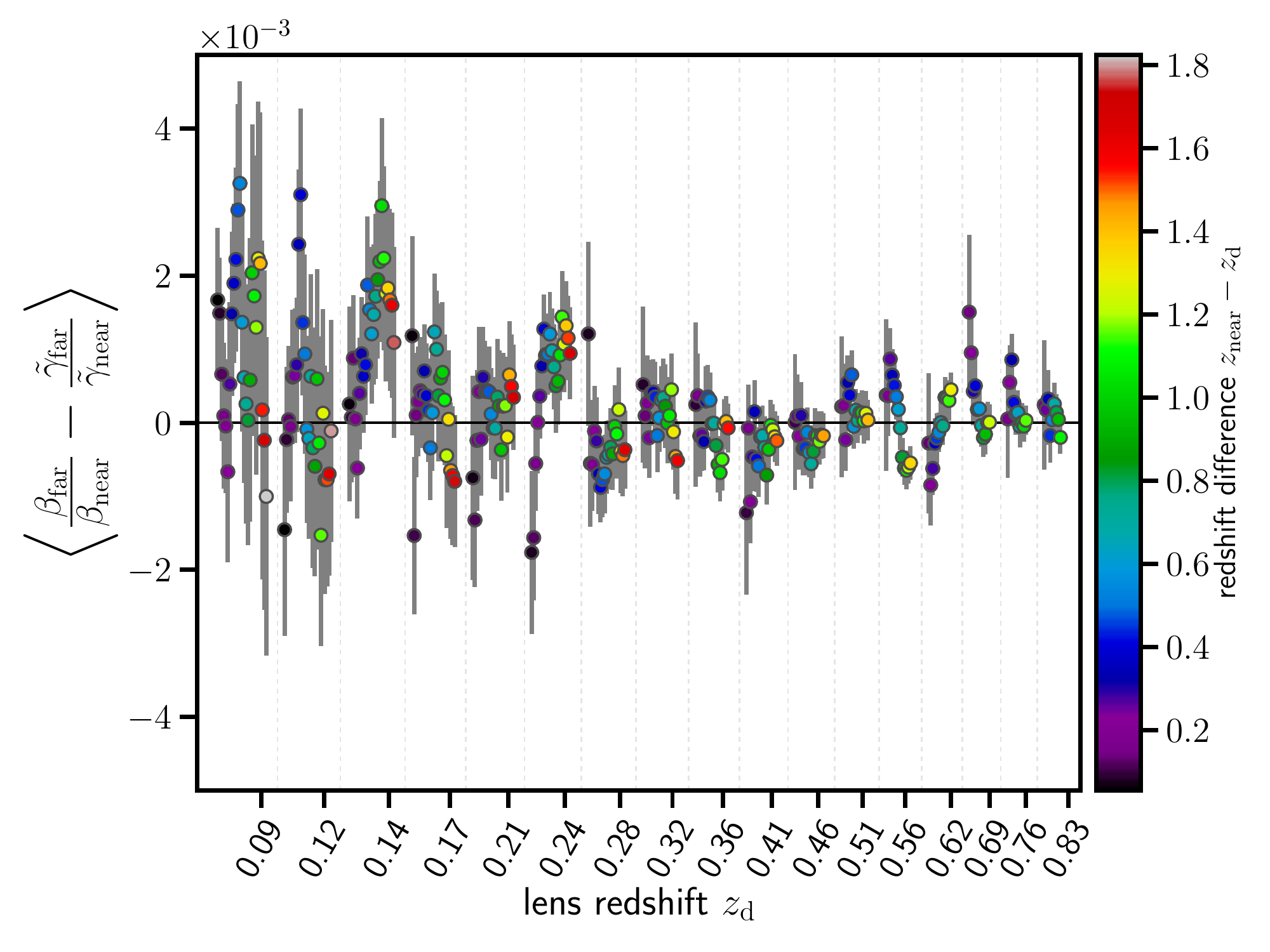}
    \caption{Shown is a shear-ratio test with the same properties as the right Fig.~\ref{pic:SRT} but with a modified shear estimator \eqref{eq:corrshearestimator} that corrects for magnification bias. While the lowest redshift bins are almost unchanged, the high redshift bins show a significant improvement.}
    \label{pic:SRTcorr}
\end{figure*}

For brevity we will introduce a new tangential shear estimator
\begin{align}
    \label{eq:corrshearestimator}
    \langle \tildegt \rangle (\zd,\zs) = \langle \hatgt \rangle (\zd,\zs) - \lambda (\zd,\zs) \langle \hatgt \rangle (\zd, \zd)\;,
\end{align}
In Fig.~\ref{pic:SRTcorr}, we show the results for the modified estimator in the realistic case of a flux-limited sample\footnote{The result for the stellar mass-limited sample are almost identical to those in Fig.~\ref{pic:SRTcorr}.}. Correcting for magnification bias indeed improves the SRT by two orders of magnitude.The scaling factor \(\lambda\) ranges from 1.1 for redshifts adjecant to the lens redshifts to \(\lambda \approx 1.7\) for \(\Delta z \approx 1\) to \(\lambda \approx 2.5\) for \(\Delta z \approx 2\). This behaviour is fairly independent of lens redshift, while \(\lambda\)-values tend to be slightly higher for low lens redshifts than for high lens redshifts.

\begin{figure}[htbp]
    \centering
    \includegraphics[width=.49\textwidth]{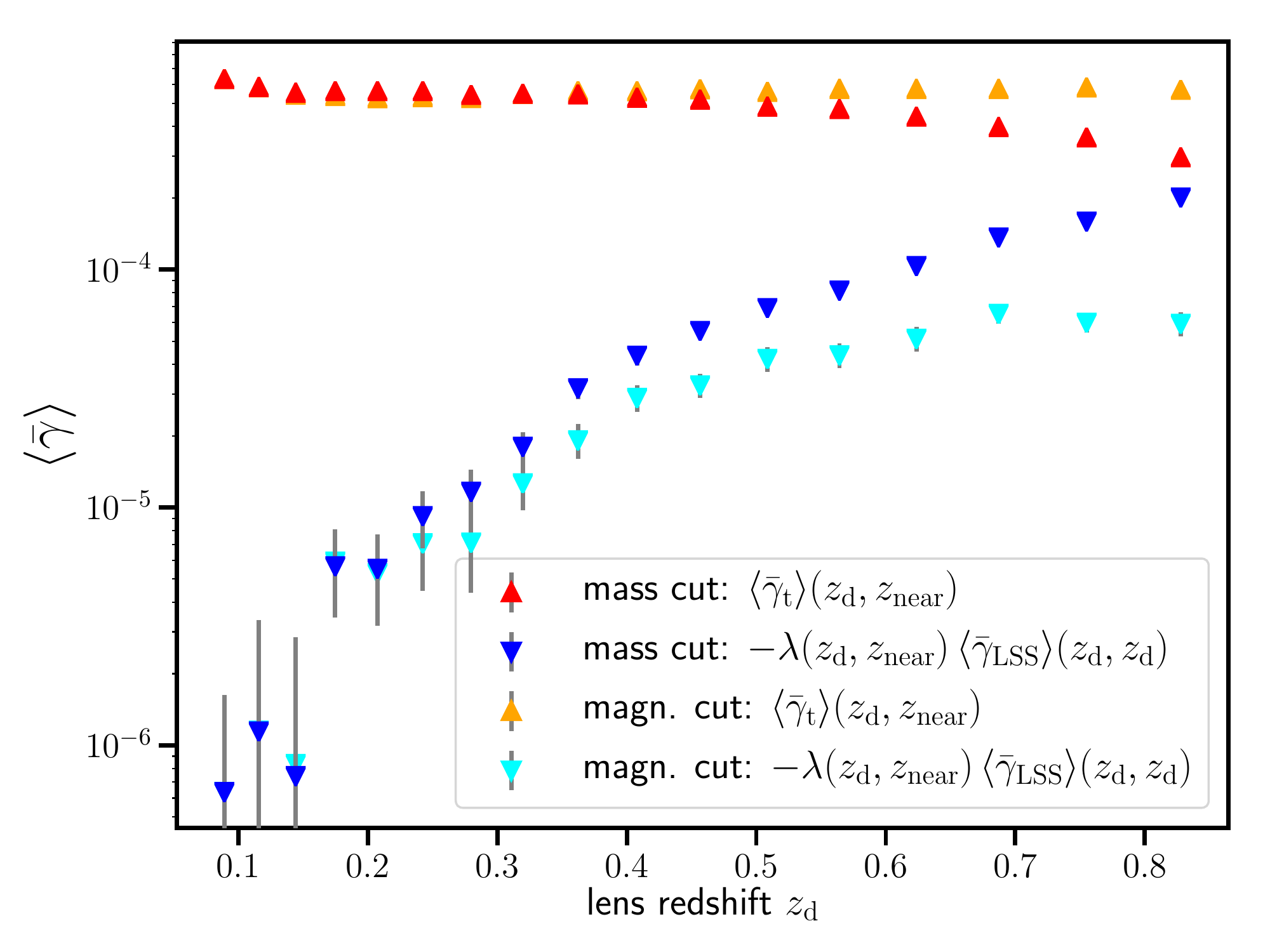}
    \caption{A comparison of \hatgt\ and \(\lambda\,\hatglss\), where we used the shear estimator \eqref{eq:shearest} within an annulus \(\theta_\mathrm{in} = 0.\!'5\) and \(\theta_\mathrm{out} = 5'\). Red and blue symbols correspond to a mass-limited sample while the orange and cyan symbols correspond to a magnitude cut \(r=24\). \(z_\mathrm{near}\) is the adjacent redshift slice of the lens redshift \zd. For low redshifts the scaling factor \(\lambda\) is around 1.2 and it decreases to 1.1 for the high redshift bins. The contribution from the LSS, \hatglss, to the shear is a steep function of lens redshift. Whereas for low redshifts the effect is negligible, the ratio of \(\hatglss/\hatgt\) approaches 1 for the highest lens redshift bin considered. The flux-limited sample is in general less affected but shows qualitatively a similar behavior.}
    \label{pic:fgdshear}
\end{figure}

To visualize the contribution from magnification effects, Fig.~\ref{pic:fgdshear} shows \(\langle\hatglss\rangle\) in comparison to the shear signal \(\langle\hatgt\rangle\) measured from the source redshift slice that is adjacent to the lens redshift slice. For low lens redshifts, \(\langle\hatglss\rangle\) makes a negligible contribution. The ratio \(\langle\hatglss\rangle/\langle\hatgt\rangle\) rises with increasing redshift. The red and blue lines correspond to a mass-limited sample and show a stronger contribution by the LSS to the shear signal than the orange and cyan lines, which represent a flux-limited sample of lens galaxies; this behavior was already seen in Fig.~\ref{pic:SRT}. The reason for this is the way magnification changes the observed number density of galaxies on the sky. For a stellar mass-limited sample the ratio of lensed galaxy number counts, \(n\), over unlensed galaxy number counts, \(n_0\), depends solely on the magnification, \(n/n_0 = \mu^{-1}\). A flux-limited sample, on the other hand, also depends on the slope of cumulative galaxy counts evaluated at the limiting flux, \( \alpha = -\, (\mathrm{d\;ln\, }n_0/\mathrm{d\; ln\, }s)\, |_{s_\mathrm{lim}} \). The ratio of lensed over unlensed galaxy number counts changes to \(n/n_0 = \mu^{\alpha-1}\).
%
%
%
%A shallow number count slope and a magnification \(\mu>1\) leads to a decrease in the local observed number density of lens galaxies, while a steep number count slope with \(\mu>1\) results in an increased number density. A flux-limited sample shows a steeper number count slope for higher lens redshifts since only the bright exponential tail of the luminosity function is visible. This is not the case for a mass-limited sample.

\subsection{Alternative ways of obtaining the scaling factor \texorpdfstring{\(\lambda\)}{lambda}}

In the previous section we showed that the scaling factor \(\lambda\) can be obtained by correlating the lensing signal at the lens redshift with that at the source redshift. We can also divide the lens population in sub-samples that show a different dependence on magnification to obtain an estimate for the scaling factor. For each of the \(N\geq2\) sub-samples Eq.~\eqref{eq:shearobs} holds
\begin{align}
    \hatgt^i (\zd, \zs) = \lambda (\zd, \zs) \hatglss^i (\zd,\zd) \, + \,\hatgg^i (\zd, \zs) \;.
\end{align}
A binning of the lenses in their magnification is, of course, dependent on magnification but unfortunately this property is not directly observable. As a proof of principle we will show the result of this approach with the magnification readily available in the ray-tracing catalogues. We already presented the dependence of shear on magnification in Fig.~\ref{pic:shearmagn} where we split the lens sample in 19 sub-samples. For a high number of lens galaxies it is sufficient to split the lens population in two samples, for example in samples with magnification bigger or smaller than the mean magnification. Then we can calculate the scaling factor \(\lambda\) with
\begin{align}
    \label{eq:altlambda}
    \lambda = \frac{\hatgt^1 - \hatgt^2}{\hatglss^1- \hatglss^2} \;,
\end{align}
where `1' corresponds to \(\mu_\mathrm{lens} < \langle \mu \rangle\) and `2' to \(\mu_\mathrm{lens} \geq \langle \mu \rangle\). Furthermore, we set \(\hatgg^1 (\zd, \zs) = \hatgg^2 (\zd, \zs)\) since the shear induced by matter correlated to the lens galaxies must be independent of the lens magnification.

We checked this approach and found similar results to our previous method for lens redshifts \(\zd > 0.15\) with a slightly worse performance for high source redshifts. For the low-redshift bins, we have only a comparatively low number of galaxies available, while at the same time the average of the LSS-induced shear \(\hatglss\) is of the order \(10^{-6}\), which is a factor of \(1000\) lower than the actual shear signal. This leads to a very noisy estimate of the scaling factor \(\lambda\).

Magnification estimates are challenging since the intrinsic variation of galaxy properties is very broad. Nonetheless, viable techniques to estimate magnification exist. One of these techniques makes use of the Fundamental Plane \citep[FP,][]{bertin2006} for elliptical galaxies. In the FP, the intrinsic effective radius of a galaxy \(R_\mathrm{eff}^\mathrm{FP}\) is related to the galaxy's surface brightness and stellar velocity  dispersion (neither of which are affected by lensing magnification). The measured effective radius \(R_\mathrm{eff}\), on the other hand, is magnified. A comparison of \(R_\mathrm{eff}\) and \(R_\mathrm{eff}^\mathrm{FP}\) estimates the magnification of a galaxy \(\mu \approx (R_\mathrm{eff}/R_\mathrm{eff}^\mathrm{FP})^2\). A modified version of the FP has been successfully applied by \citet{huff2011} to the photometric catalogue from the Sloan Digital Sky Survey (SDSS).

The intrinsic scatter in the FP is \(\sim 20\%\) \citep{bernardi2003} which corresponds to a \(\sigma_\mu \sim 40\%\) scatter in the magnification estimate. We repeat our analysis with a scaling factor as measured in Eq.~\eqref{eq:altlambda} but now add Gaussian noise with mean zero and width \(\sigma_\mu\) to the magnification from the ray-tracing catalogue. Fortunately, the introduced scatter does not invalidate the mitigation for lens redshifts \(\zd > 0.2\). The reason for this is the almost linear relation between magnification and shear as seen in Fig.~\ref{pic:fgdshear}. For lenses with \(\zd \le 0.2\) the noise in the magnification estimate enhances the variance in the SRT by a factor of \(\sim 20\). This is again due to the very small foreground signal (compare Fig.~\ref{pic:fgdshear}). The difference in the foreground signal \hatglss \ between bin `1' and `2' is so small that the introduced scatter in magnification can bring the difference very close to zero and can thus lead to unreasonably high values in the scaling factor. The enhanced scatter in the SRT can be suppressed by enforcing the scaling factor to be smaller than 5, which is physically justified. With the added constraint on \(\lambda\) the alternative mitigation strategy for the magnification bias performs nearly as well as shown in Fig.~\ref{pic:SRTcorr}.

\subsection{Impact of shape noise}

In this work we made use of ray-tracing simulation that have only minor contributions of noise. In observations, the largest source of uncertainty in weak lensing measurements is shape noise \citep[e.g.,][]{niemi2015}. It arises because the measured galaxy ellipticities are dominated by the intrinsic galaxy shapes, with a much weaker contribution from lensing shear. The intrinsic ellipticity is expected to be randomly distributed and thus, the average over a sufficiently large number of galaxies vanishes. We employ a simplified model of additive, uncorrelated Gaussian noise to obtain an estimate of how shape noise affects our mitigation strategy. We add a Gaussian with zero mean and width \(\sigma_\epsilon = 0.3\) to each Cartesian shear component from the ray-tracing catalogue before estimating the tangential shear around lens galaxies. Furthermore, we limit the density of source galaxies to \(<35\,\mathrm{gal/arcmin}^2\) which is certainly fulfilled with a magnitude cut in the \(r\)-band at \(24\,\mathrm{mag}\). The last adjustment we make for the limiting magnitude for the lens galaxies, it is reduced to \(22\,\mathrm{mag}\) which leaves \(\sim 1/3\) of the lens galaxies with a \(24\,\mathrm{mag}\)-cut. Since we reduced the density of background galaxies by a factor of \(\sim 2500\), the SRT shows a way more noisy result. For some combinations of lens and source redshifts the corresponding variance increased by a factor of 10\,000. As one of the reasons, we identify the decreased statistics, however, we must also take into account how we constructed the null hypothesis of the SRT (\ref{eq:nullhyp}). It contains a ratio of shear values, nominator as well as denominator are noisy quantities and if the denominator is close to zero, we obtain very high values in the SRT. Hence, we propose to change the null hypothesis to
\begin{align}
    \label{eq:newnullhypo}
    \langle \hatgg (\zd, z_j) \rangle \, \beta(\zd,z_i) - \langle \hatgg ( \zd, z_i) \rangle\, \beta(\zd,z_j) = 0\;.
\end{align}
With the new estimator the noise increased for all combinations of lens and source redshifts roughly by a factor 100 without extreme outliers as before.

To still verify that our mitigation strategy improves the SRT, we perform a reduced \(\chi^2\) test and the result is shown in table \ref{tab:chitest}.
\begin{table}[]
    \caption{Results of a reduced \(\chi^2\) test are shown. We perform the test for comparability for all cases with the modified null hypothesis (\ref{eq:newnullhypo}). The first two rows correspond to a mass-limited sample, the next two to a magnitude limited sample and the last two consider also shape noise. For each we show the \(\chi^2_\mathrm{red}\) result with and without mitigation. As can be seen by eye, the mitigation drastically improves the result in the first two cases. In the case that include noise the \(\chi^2_\mathrm{red}\) test still performs better when the mitigation is used.}
    \label{tab:chitest}
    \centering
    \begin{tabular}{cccr}
        \hline\hline
        lens mag limit & source mag limit & mitigation & \(\chi^2_\mathrm{red}\) \\
        \hline
        none  & none    & no  & 80.90 \\
        none  & none    & yes & 1.06 \\
        24mag & none    & no  & 10.97 \\
        24mag & none    & yes & 1.02 \\
        22mag & 24.5mag & no  & 2.28 \\
        22mag & 24.5mag & yes & 1.36 \\
        \hline
    \end{tabular}
\end{table}
As expected, the mitigation improves the \(\chi^2_\mathrm{red}\) value significantly for the cases shown in Fig.~\ref{pic:SRT}. For the shape noise dominated case a SRT with corrected magnification bias still performs better than without mitigation, although the difference is less pronounced. It is important to note, that the level of shape noise reduces with increased observed area on the sky. On the other hand, the magnification bias is fairly independent of observed area. Thus, future surveys with observed areas \(\gtrsim 1000\,\mathrm{deg}^2\) will have less impact of shape noise than we considered here (and vice versa).

\subsection{Impact of intrinsic alignments}

So far we ignored that the intrinsic ellipticities of galaxies may be correlated with the positions of other galaxies that are close in real space \citep[e.g.][]{Joachimi2015}. This flavor of intrinsic alignment (IA) may impact our estimate of the LSS contribution to the shear signal at the lens redshift. In particular, Eq.~\eqref{eq:estimator_for_hatglss_of_z_d_z_d} is modified
\begin{equation}
\label{eq:estimator_for_hatglss_of_z_d_z_d_with_IA}
\langle \hatgt \rangle(\zd, \zd) 	= \langle \hatglss \rangle(\zd,\zd) +   \langle \hatgia \rangle(\zd,\zd).
\end{equation}
The IA contribution is
\begin{equation}
\label{eq:hatgia}
\langle \hatgia \rangle(\zd,\zd) =
 \int_{\theta_{\mathrm{in}}}^{\theta_{\mathrm{out}}} 2\pi\,\theta\,\mathrm{d}\theta \, \langle\gamma_\mathrm{IA}\rangle (\theta) \, \frac{1}{2\pi \theta^2}.
\end{equation}
The intrinsic tangential ellipticity $\langle\gamma_\mathrm{IA}\rangle (\theta)$ as a function of angular separation can be roughly estimated by:
\begin{equation}
\label{eq:gia}
 \langle\gamma_\mathrm{IA}\rangle (\theta) \approx \frac{\wdI(\theta \chid, \zd)}{\Delta\chi(\Delta z) + \wdd(\theta \chid, \zd)}.
\end{equation}
Here, $\wdI(r,z)$ denotes the projected cross correlation at transverse comoving separation $r$ and redshift $z$ between the lens galaxy positions and the tangential components of the source galaxy intrinsic ellipticities, $\wdd(r,z)$ denotes the projected cross correlation of the lens and source galaxy positions, and $\Delta\chi$ is the projection depth corresponding to the redshift interval $\Delta z$ used to select source galaxies around the lens redshift.

\begin{figure}
\centerline{\includegraphics[width=\linewidth]{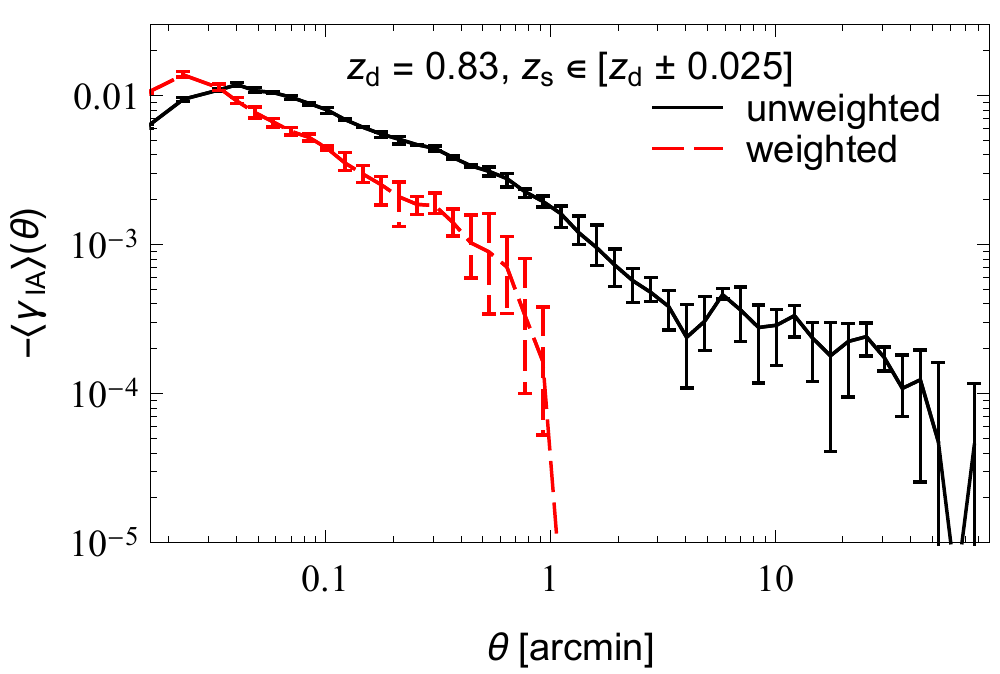}}
\caption{
\label{pic:gia}
The mean intrinsic tangential ellipticity $\langle\gamma_\mathrm{IA}\rangle (\theta)$ as a function of angular separation $\theta$ 
\citep[estimated from the Illustris simulation,][]{Vogelsberger2014, Hilbert2017}
for lenses at redshift $\zd = 0.83$ and sources at redshifts $\zs \in [0.805, 0.855]$. 
The IA signal strongly depends on whether the shape estimator uses unweighted or radially weighted second moments of the galaxy image light distribution.
}
\end{figure}

 We estimate the IA contribution using measurements of the projected IA correlations by \cite{Hilbert2017} in the Illustris simulation \citep[][]{Vogelsberger2014}. The correlation $\wdI$, and thus $\langle \hatgia \rangle(\zd,\zd)$, depend on the source and lens galaxy sample selection criteria, and also on how much weight the galaxy image shear estimator gives to the outskirts of the galaxy images.

The resulting $\langle\gamma_\mathrm{IA}\rangle (\theta)$ is shown in Fig.~\ref{pic:gia} for lenses at $\zd = 0.83$ and sources between $\zs = \zd - \Delta z/2$ and $\zs = \zd + \Delta z/2$, where $\Delta z = 0.05$. This yields $\langle \hatgia \rangle(\zd,\zd) \approx -3\times 10^{-4}$ for radially weighted moments \citep[such as][KSB]{Kaiser1995}, which is noticeable compared to the LSS contribution $\langle \hatglss \rangle(\zd,\zd)$ (see Fig.~\ref{pic:fgdshear}).

Equation~\eqref{eq:gia} shows that the IA contribution $\langle \hatgia \rangle(\zd,\zd)$ could be reduced by substantially increasing the source redshift window size $\Delta z$.
Furthermore, IA correlations appear to be dominated by galaxies in the same halo and do not reach beyond a few tens of Mpc.
Thus, the IA contribution could be avoided if one can reliably select source-lens pairs such that the sources are at least a few tens of Mpc in front of the lenses. 
Moreover, the IA contribution $\langle \hatgia \rangle(\zd,\zd)$ can be substantially reduced by increasing the lower integration bound $\theta_{\mathrm{in}}$. For example, $\langle \hatgia \rangle$ practically vanishes for KSB-like estimators and $\theta_{\mathrm{in}} \geq 1\,\arcmint$.

\section{\label{Sc6}Discussion \& Conclusion}

In this article we take a closer look at the influence of magnification bias on the shear-ratio test as introduced in \citet{jain2003} as well as a viable mitigation strategy. An advantage of the SRT is that it can be applied to the same data as obtained from cosmic shear surveys. Moreover, it is a purely geometrical method and does not rely on any assumptions of structure growth. As such, this null test has the potential to uncover remaining systematics in shear measurement and redshift estimation. \citet{schneider2016} even extended the SRT in such a way that it does not depend on the choice of cosmology anymore.

The SRT is based on a ratio of shears that is induced by matter correlated with the lens galaxies, and it does not take LSS effects into account. LSS mainly changes the observed shear signal due to magnification which alters the observed number density of lens galaxies on the sky.
%Suppose a massive object like a galaxy cluster is situated between us and some distant sources, then due to stretching of the apparent solid angle on the sky the observed number density of the galaxies decreases. However, since faint galaxies show also a magnified flux, we will detect objects that would have been to faint otherwise; thus, increasing the number density on the sky. The net outcome is determined by the number count slope. 
\citet{hilbert2009} showed that, depending on the observed scale, the tangential shear can deviate by up to \(20\%\) from  the GGL signal expected from shear that is induced by matter correlated to the lens galaxies, though this fact seems to have been largely ignored in subsequent observational studies. By using shear power spectra, \citet{ziour2008} derived relations that suggested that magnification bias influences the SRT quite heavily.

We made use of ray-tracing results through the Millennium simulation \citep{hilbert2009} and galaxy catalogues from semi-analytic models \citep{henriques2015} to obtain accurate estimates for the tangential shear around galaxies at several redshifts. We used lenses in the redshift range \(0.09 \leq \zd \leq 0.83\) and sources in the range \(\zd < \zs \leq 1.9\). With that we were able to quantify the impact of magnification bias on the SRT as can be seen in Fig.~\ref{pic:SRT} and \ref{pic:devSRT}. We find: (1) the higher the lens redshift, the larger is the deviation of the SRT from its expected value -- for the lens redshifts considered the deviation from 0 increases from \(10^{-3}\) by a factor of \(\sim\)100, (2) lenses and sources must be well separated along the line-of-sight -- the relative impact of the magnification bias on the SRT is largest when the source and lens galaxies are close, and (3) magnification bias depends on the range over which shear is estimated.

For our mitigation strategy we assume that the shear signal is a superposition of the shear induced by matter correlated with the lens galaxies at redshift \zd \ and shear due to matter between us and the source galaxies, where the LSS in the redshift range \(\zd < z< \zs\) is irrelevant for the magnification bias on the shear signal. To extract the LSS-induced shear signal from the data, we measure the tangential shear around lens galaxies at the lens redshift and use a scaling factor to approximate its value at the source redshift. The scaling factor can be calculated as a correlation between these two shear components (Eq.~\ref{eq:fudgefac}). Subtracting the scaled LSS-induced shear from the measured shear signal will yield the shear that is induced by matter correlated with the lens galaxies. The latter is what is needed for the SRT, and the good performance of this mitigation approach can be seen in Fig.~\ref{pic:SRTcorr}. We further introduced an alternative way of obtaining the scaling factor that relies on dividing the lens population into sub-samples with different magnification. Estimating the magnification with the Fundamental Plane for early-type galaxies leads to results that perform well for all redshifts where magnification bias is important. Furthermore, we estimated the impact of shape noise on our mitigation strategy. All redshifts show an increased scatter by a factor of \(\sim 100\). A \(\chi^2_\mathrm{red}\) analysis showed that applying the mitigation still improves the SRT. Also, shape noise can be reduced by observing a larger area on the sky in contrast to the magnification bias. We used roughly \(1000\,\mathrm{deg}^2\) for our analysis, future experiments like Euclid will surpass this by a factor of \(\ge 10\). Finally, we discussed the possible impact of intrinsic alignments (IA) on our mitigation strategy. Since the IA contribution to the shear signal at the lens redshift might be substantial compared to the LSS contribution, modifications to our mitigation strategy that reduce the impact of IA (e.g. by estimating the LSS contribution using sources slightly in front of the lenses) should be explored in more detail in future work.

Magnification bias is present on all relevant scales and needs to be corrected for. It affects not only the performance of the SRT, but must be considered in all applications of GGL and its generalization to groups and clusters. A viable mitigation strategy is therefore crucial for ongoing and future experiments.

\begin{acknowledgements}
    %We would like  for valuable discussions,for valuable comments on this paper, and  for his constructive comments and advice. 
    Part of this work was supported by the German \emph{Deut\-sche For\-schungs\-ge\-mein\-schaft, DFG\/} project numbers SL\,172/1-1 and SCHN\,342/13-1.
    Sandra Unruh is a member of the International Max Planck Research School (IMPRS) for Astronomy and Astrophysics at the Universities of Bonn and Cologne. Stefan Hilbert acknowledges support by the DFG cluster of excellence \lq{}Origin and Structure of the Universe\rq{} (\href{http://www.universe-cluster.de}{\texttt{www.universe-cluster.de}}).
\end{acknowledgements}

\bibliography{Magnbiasbib}
\bibliographystyle{aa}

\end{document}